\documentclass[prc,preprint,showpacs,preprintnumbers,amsmath,amssymb]{revtex4}

\usepackage[mathscr]{eucal}
\usepackage{graphicx}
\def\slr#1{\setbox0=\hbox{$#1$}           
   \dimen0=\wd0                                 
   \setbox1=\hbox{/} \dimen1=\wd1               
   \ifdim\dimen0>\dimen1                        
      \rlap{\hbox to \dimen0{\hfil/\hfil}}      
      #1                                        
   \else                                        
      \rlap{\hbox to \dimen1{\hfil$#1$\hfil}}   
      /                                         
   \fi}

\def\kp{k^{\,\prime}}
\def\kpsq{k^{\,\prime\,2}}
\def\ksq{k^2}

\def\myint#1{\!\int\!\!\frac{d^4\!{#1}}{(2\pi)^4}\,}
\def\mytint#1{\!\int\!\!\frac{d^3\!{#1}}{(2\pi)^3}\,}
\def\gev#1{ GeV${}^{#1}$}
\def\be{\begin{eqnarray}}
\def\ee{\end{eqnarray}}

\renewcommand{\theequation}%
    {\arabic{section}.\arabic{equation}}
\makeatletter \@addtoreset{equation}{section} \makeatother

\begin{document}


\title{Chiral Symmetry Restoration and Deconfinement of Light Mesons at Finite Temperature}

\author{Hu Li}
\author{C. M. Shakin}
 \email[email:]{casbc@cunyvm.cuny.edu}

\affiliation{%
Department of Physics and Center for Nuclear Theory\\
Brooklyn College of the City University of New York\\
Brooklyn, New York 11210
}%

\date{August, 2002}

\begin{abstract}
There has been a great deal of interest in understanding the
properties of quantum chromodynamics (QCD) for a finite value of
the chemical potential and for finite temperature. Studies have
been made of the restoration of chiral symmetry in matter and at
finite temperature. The phenomenon of deconfinement is also of
great interest, with studies of the temperature dependence of the
confining interaction reported recently. In the present work we
study the change of the properties of light mesons as the
temperature is increased. For this study we make use of a
Nambu--Jona-Lasinio (NJL) model that has been generalized to
include a covariant model of confinement. The parameters of the
confining interaction are made temperature-dependent to take into
account what has been learned in lattice simulations of QCD at
finite temperature. The constituent quark masses are calculated at
finite temperature using the NJL model. A novel feature of our
work is the introduction of a temperature dependence of the NJL
interaction parameters. (This is a purely phenomenological feature
of our model, which we do not attempt to derive from more
fundamental considerations.) With the three temperature-dependent
aspects of the model mentioned above, we find that the mesons we
study are no longer bound when the temperature reaches the
critical temperature, $T_c$, which we take to be 170 MeV. We
believe that ours is the first model that is able to describe the
interplay of chiral symmetry restoration and deconfinement for
mesons at finite temperature. The introduction of
temperature-dependent coupling constants is a feature of our work
whose further consequences should be explored in future work.
\end{abstract}


\maketitle

\section{INTRODUCTION}

In recent years we have studied a generalized version of the
Nambu--Jona-Lasinio (NJL) model which includes a covariant model
of Lorentz-vector confinement [1-5]. Extensive applications have
been made in the study of light mesons, with particularly
satisfactory results for the properties of the $\eta (547)$,
$\eta^\prime (958)$ and their radial excitations [6]. Since the
modifications of the confining potential at finite temperature
have recently been obtained in lattice simulations of QCD [7-12]
(see Fig. 1), we became interested in introducing that feature in
our generalized NJL model, whose Lagrangian is \be {\cal L}=&&\bar
q(i\slr
\partial-m^0)q +\frac{G_S}{2}\sum_{i=0}^8[
(\bar q\lambda^iq)^2+(\bar qi\gamma_5 \lambda^iq)^2]\nonumber\\
&&-\frac{G_V}{2}\sum_{i=0}^8[
(\bar q\lambda^i\gamma_\mu q)^2+(\bar q\lambda^i\gamma_5 \gamma_\mu q)^2]\nonumber\\
&& +\frac{G_D}{2}\{\det[\bar q(1+\gamma_5)q]+\det[\bar
q(1-\gamma_5)q]\} \nonumber\\
&&+ {\cal L}_{conf}\,. \ee Here, the $\lambda^i(i=0,\cdots, 8)$
are the Gell-Mann matrices, with
$\lambda^0=\sqrt{2/3}\mathbf{\,1}$, $m^0=\mbox{diag}\,(m_u^0,
m_d^0, m_s^0)$ is a matrix of current quark masses and ${\cal
L}_{conf}$ denotes our model of confinement.

 \begin{figure}
 \includegraphics[bb=0 0 250 350, angle=-90, scale=1.2]{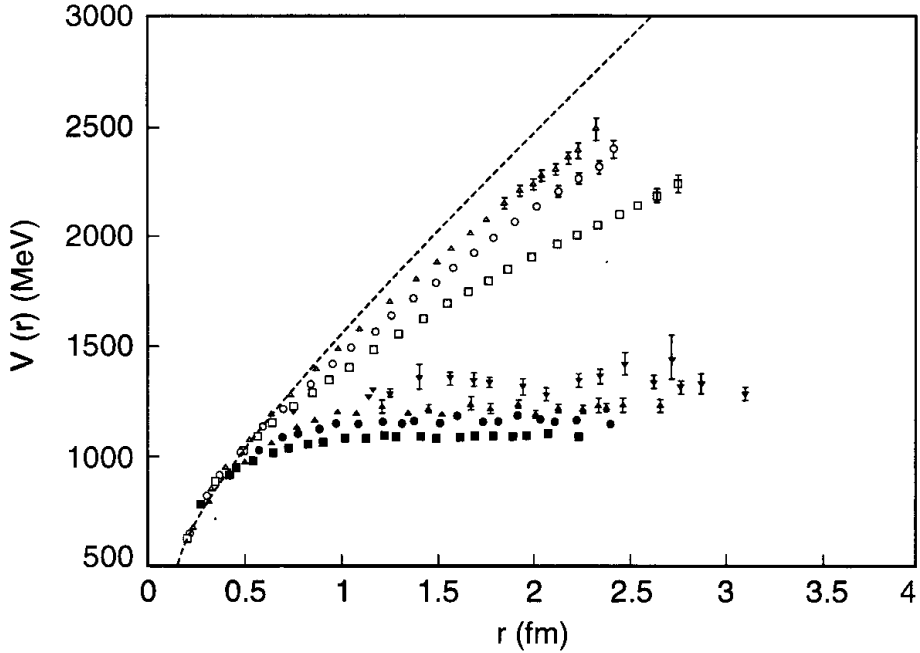}%
 \caption{A comparison of quenched (open symbols) and unquenched (filled symbols) results for
 the interquark potential at finite temperature [7]. The dotted line is the zero temperature
 quenched potential. Here, the symbols for $T=0.80T_c$ [open triangle], $T=0.88T_c$
 [open circle], $T=0.94T_c$ [open square], represent the quenched
 results. The results with dynamical fermions are given at $T=0.68T_c$ [solid downward-pointing
 triangle], $T=0.80T_c$ [solid upward-pointing triangle], $T=0.88T_c$ [solid circle],
 and $T=0.94T_c$ [solid square].}
 \end{figure}

The dependence of the constituent masses of the NJL model upon the
density of nuclear matter has been studied in earlier work, where
we also studied the deconfinement of light mesons with increasing
matter density [13]. In that work we introduced some density
dependence of the coupling constants of the model. However, that
was not done in a systematic fashion. As we will see, in our study
of the temperature dependence of the constituent masses, as well
as in the calculation of temperature-dependent hadron masses, we
introduce a temperature dependence of the coupling constants,
which for $T=T_c$, represents a 17\% reduction of the magnitude of
these constants. Since the use of temperature-dependent coupling
constants is a novel feature of our work, we provide some evidence
that such temperature dependence is necessary to create a
formalism that is consistent with what is known concerning QCD
thermodynamics. This aspect of our work is discussed in the
Appendix.

The organization of our work is as follows. In Section II we
review our treatment of confinement in our generalized NJL model
and describe the modification we introduce to specify the
temperature dependence of our confining interaction. In Section
III we discuss the calculation of the temperature dependence of
the constituent masses of the up, down and strange quarks. In
Section IV we describe how random-phase-approximation (RPA)
calculations may be made to obtain the wave function amplitudes
and masses of the light mesons. (We consider the properties of the
$\pi$, $K$, $f_0$, $a_0$ and $K_0^*$ mesons and their radial
excitations in this work.) In Section V we provide the results of
our calculations of the meson spectra at finite temperature.
Finally, Section VI contains some further discussion and
conclusions.

\section{covariant lorentz-vector model of confinement}

We have published a number of papers in which we have described
our model of confinement[1-6]. In this work we provide a review of
the important features of the model. As a first step, we introduce
a vertex function for the confining interaction. [See Fig. 2.] For
example, the vertex useful in the study of pseudoscalar mesons
satisfies an equation of the form [6] \be
\overline\Gamma_{5,ab}(P,k)&=&\gamma_5-i\myint\kp[\gamma^\mu
S_a(P/2+\kp)\\\nonumber&&\times\overline\Gamma_{5,ab}(P,\kp)S_b(-P/2+\kp)\gamma_\mu]
V^C(\vec k-\vec k^\prime)\,,\ee where $S_a(P/2+\kp)=[\slr P/2+\slr
k\,^\prime-m_a+i\eta]^{-1}$. Here, $m_a$ and $m_b$ are the
constituent quark masses. We consider the form $V^C(r)=\kappa
r\exp[-\mu r]$ and obtain the Fourier transform \be V^C(\vec
k-\vec k\,^\prime)=-8\pi\kappa\left[\frac1{[(\vec k-\vec
k\,^\prime)^2+\mu^2]^2}-\frac{4\mu^2}{[(\vec k-\vec
k\,^\prime)^2+\mu^2]^3}\right]\,,\ee which is used in Eq. (2.1).
Note that the matrix form of the confining interaction is
$\overline V{}\,^C(\vec k-\vec k\,^\prime)=\gamma^\mu(1)V^C(\vec
k-\vec k\,^\prime)\gamma_\mu(2)$, since we are using
Lorentz-vector confinement. The form of the potential may be made
covariant by introducing the four-vectors  \be \hat
k^\mu=k^\mu-\frac{(k\cdot P)P^\mu}{P^2}\,, \ee  and \be \hat
k^{\prime\,\mu}=k^{\prime\,\mu}-\frac{(k^\prime\cdot
P)P^\mu}{P^2}\,. \ee We then define \be V^C(\hat k-\hat
k\,^\prime)=-8\pi\kappa\left[\frac1{[-(\hat k-\hat
k\,^\prime)^2+\mu^2]^2}-\frac{4\mu^2}{[-(\hat k-\hat
k\,^\prime)^2+\mu^2]^3}\right]\,,\ee which reduces to $V^C(\vec
k-\vec k\,^\prime)$ of Eq. (2.2) in the meson rest frame, where
$\vec P=0$.

 \begin{figure}
 \includegraphics[bb=-30 0 150 600, angle=90, scale=0.5]{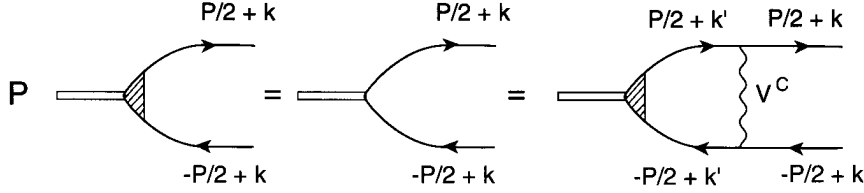}%
 \caption{The figure provides a schematic representation of Eq. (2.1)
 for the confining vertex function $\overline\Gamma_{5,ab}(P,k)$.}
 \end{figure}

It is also useful to define scalar functions
$\Gamma_{5,ab}^{+-}(P,k)$ and $\Gamma_{5,ab}^{-+}(P,k)$ [6]. We
introduce \be \Lambda_a^{(+)}(\vec k)=\frac{\slr k_a+m_a}{2m_a}\ee
and \be \Lambda_a^{(-)}(-\vec k)=\frac{\slr {\widetilde
k}_a+m_a}{2m_a}\,,\ee where $k^\mu=[E_a(\vec k), \vec k]$ and
$\widetilde k^\mu=[-E_a(\vec k), \vec k]$, and define \be
\Lambda_a^{(+)}(\vec k)
\overline\Gamma_{5,ab}(P,k)\Lambda_b^{(-)}(-\vec
k)=\Gamma_{5,ab}^{+-}(P,k)\Lambda_a^{(+)}(\vec
k)\gamma_5\Lambda_b^{(-)}(-\vec k)\,,\ee and \be
\Lambda_a^{(-)}(-\vec k)
\overline\Gamma_{5,ab}(P,k)\Lambda_b^{(+)}(\vec
k)=\Gamma_{5,ab}^{-+}(P,k)\Lambda_a^{(-)}(-\vec
k)\gamma_5\Lambda_b^{(+)}(\vec k)\,.\ee We have obtained equations
for $\Gamma_{5}^{+-}(P,k)$ and $\Gamma_{5}^{-+}(P,k)$. For
example, with $m_a=m_b$, we have \be \Gamma_{5}^{+-}(P^0,|\vec
k|)=1-\mytint\kp\left[\frac{m^2-2E(\vec k)E(\vec
k\,^\prime)}{E(\vec k)E(\vec k\,^\prime)}\right]
\frac{\Gamma_{5}^{+-}(P^0,|\vec k\,^\prime|)V^C(\vec k-\vec
k\,^\prime)}{P^0-2E(\vec k\,^\prime)}\,.\ee Note that
$\Gamma_{5}^{+-}(P^0,|\vec k\,^\prime|)=0$, when $P^0=2E(\vec
k\,^\prime)$, so that one need not introduce a small quantity,
$i\epsilon$, in the denominator on the right-hand side of Eq.
(2.10). (Alternatively, we may note that
$\Gamma_{5}^{+-}(P^0,|\vec k|)=0$ when the quark and antiquark in
Fig. 2 are on mass shell.) The confinement vertex functions
defined in our work may be used to calculate vacuum polarization
functions which are real functions. The unitarity cut, that would
otherwise be present, is eliminated by the vertex functions which
vanish when both the quark and antiquark go on mass shell [1-6].

We have presented Eq. (2.10), since we wish to stress that for the
study of light mesons the constituent quark mass, $m$, is of the
order of $|\vec k|$, so that $|\vec k|/m$ is not small. Therefore,
the term in the square bracket on the right-hand side of Eq.
(2.10), which would be equal to $-1$ in the nonrelativistic limit,
provides quite important relativistic corrections to the form
given in Eq. (2.5).

In Fig. 3a we show the homogeneous equation for the confining
vertex. The solution of the homogeneous equation allows us to
construct the wave functions bound in the confining field. For
example, we may define \be
\psi_i^{(+)}(k)=\frac{\Gamma_{5}^{+-}(P_i^0, k)}{P_i^0-2E(\vec
k)}\ee and \be \psi_i^{(-)}(k)=-\frac{\Gamma_{5}^{-+}(-P_i^0,
k)}{P_i^0+2E(\vec k)}\,,\ee which we may term the ``large" and
``small" components of the wave function of the bound state with
energy $P_i^0$. In Fig. 3b we show the equation for the vertex
function that includes both the effects of the short-range NJL
interaction and the confining interaction. We will return to a
consideration of the equations obtained in an analysis of Fig. 3b
in Section IV where we consider the RPA equations of our model.

 \begin{figure}
 \includegraphics[bb=30 0 180 220, angle=90, scale=1.5]{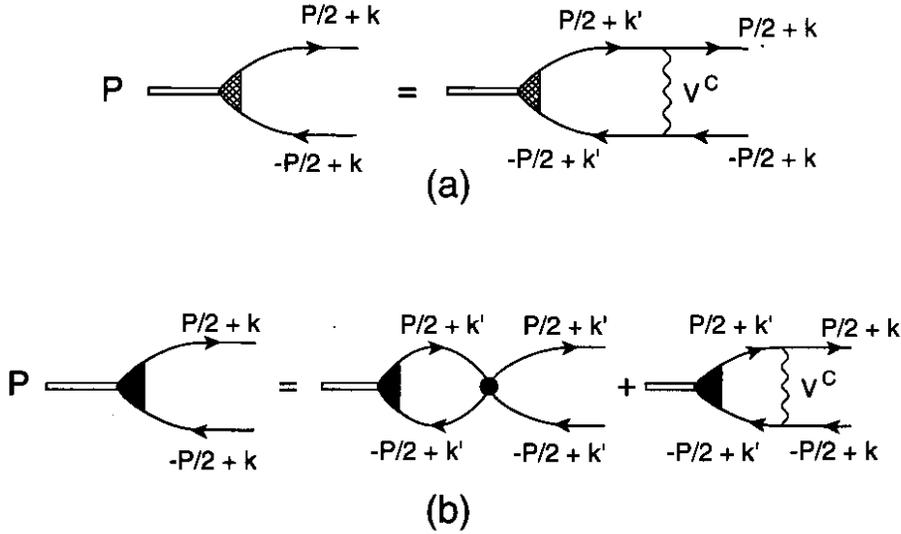}%
 \caption{a) Bound states in the confining field (wavy line) may be found by solving
 the equation for the vertex shown in this figure, b) Effects of both the confining
 field and the short-range NJL interaction (filled circle) are included when solving
 for the vertex shown in this figure.}
 \end{figure}

In order to motivate our treatment of the temperature dependence
of the confining interaction, we have presented some results
obtained with dynamical quarks (filled symbols) in Fig. 1. The
fact that the potential becomes (approximately) constant for $r>1$
fm is ascribed to ``string breaking" in the presence of dynamical
quarks. (Note that, upon string breaking, the force between the
infinitely massive quark and antiquark vanishes.)

For our calculations we have used $\mu=0.010$ GeV and
$\kappa=0.055$\gev2 in the past. In order to introduce temperature
dependence in our model, we replace $V^C(r)=\kappa r\exp[-\mu r]$
by $V^C(r, T)=\kappa r\exp[-\mu(T) r]$, with \be
\mu(T)=\frac{\mu_0}{1-0.7(T/T_c)^2}\,,\ee where $\mu_0=0.01$ GeV.
Values of $V^C(r,T)$ for various values of the ratio $T/T_c$ are
given in Fig. 4. We remark that, while $V^C(r)\rightarrow 0$ for
large $r$, the bound-state solutions found for $V^C(r)$ are
largely unaffected, since barrier penetration effects are
extremely small in our model. The maximum value of the potential
is $V_{max}^C=\kappa/\mu e$ with the corresponding value of
$r_{max}=1/\mu$. Thus, in the study of the bound states, our model
is essentially equivalent to one with $V^C(r)=\kappa r\exp[-\mu
r]$ for $r<r_{max}$ and $V^C(r)=V_{max}^C$ for $r>r_{max}$. The
same remarks pertain, if we replace $\mu$ by $\mu(T)$ of Eq.
(2.13). With that replacement, we have \be
V_{max}(T)=\frac{\kappa[1-0.7(T/T_c)^2]}{\mu_0e}\,.\ee We note
that $V_{max}$ is finite at $T=T_c$ , a result that is in general
accord with what is found in lattice calculations of the
interquark potential for massive quarks.

 \begin{figure}
 \includegraphics[bb=0 0 300 200, angle=0, scale=1.2]{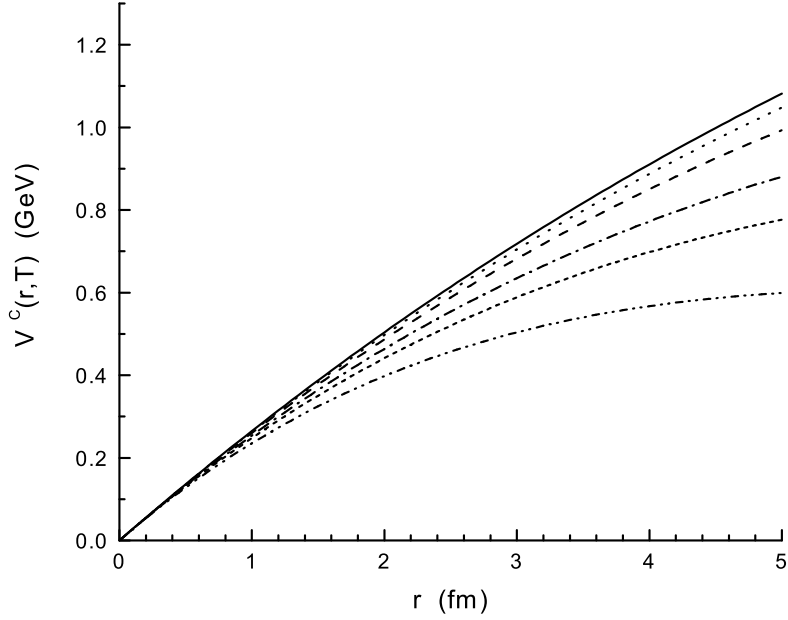}%
 \caption{The potential $V^C(r, T)$ is shown for $T/T_c=0$ [solid line],
 $T/T_c=0.4$ [dotted line], $T/T_c=0.6$ [dashed line], $T/T_c=0.8$ [dash-dot line],
 $T/T_c=0.9$ [short dashes], $T/T_c=1.0$ [dash-dot-dot line]. Here,
 $V^C(r,T)=\kappa r\exp[-\mu(T)r]$, with $\mu(T)=0.01\mbox{GeV}/[1-0.7(T/T_c)^2]$.}
 \end{figure}

\section{calculation of constituent quark masses at finite temperature}

In an earlier work we carried out a Euclidean-space calculation of
the up, down, and strange constituent quark masses taking into
account the 't Hooft interaction and our confining interaction
[14]. The 't Hooft interaction plays only a minor role in the
coupling of the equations for the constituent masses. If we
neglect the confining interaction and the 't Hooft interaction in
the mean field calculation of the constituent masses, we can
compensate for their absence by making a modest change in the
value of $G_S$. [See Eq. (1.1).] For the calculations of this work
we calculate the meson masses using the formalism presented in the
Klevansky review [15]. (Note that our value of $G_S$ is twice the
value of $G$ used in that review.) The relevant equation is Eq.
(5.38) of Ref. [15]. Here, we put $\mu=0$ and write \be
m(T)=m^0+4GN_C\frac{m(T)}{\pi^2}\int_0^\Lambda
dp\frac{p^2}{E_p}\tanh(\frac 1 2\beta E_p)\,,\ee where
$\Lambda=0.631$ GeV is a cutoff for the momentum integral,
$\beta=1/T$ and $E_p=[\vec p^2+m^2(T)]^{1/2}$. In our calculations
we replace $G$ by $G_S(T)/2$ and solve the equation \be
m(T)=m^0+2G_S(T)N_C\frac{m(T)}{\pi^2}\int_0^\Lambda
dp\frac{p^2}{E_p}\tanh(\frac 1 2\beta E_p)\,,\ee with
$G_S(T)=11.38[1-0.17\,T/T_c]$ GeV, $m_u^0=0.0055$ GeV and
$m_s^0=0.120$ GeV. Thus, we see that $G_S(T)$ is reduced from the
value $G_S(0)$ by 17\% when $T=T_c$. The results obtained in this
manner for $m_u(T)$ and $m_s(T)$ are shown in Fig. 5. Here, the
temperature dependence we have introduced for $G_S(T)$ serves to
provide a somewhat more rapid restoration of chiral symmetry than
that which is found for a constant value of $G_S$. That feature
and the temperature dependence of the confining field leads to the
deconfinement of the light mesons considered here at $T=T_c$. (See
Section V.)

 \begin{figure}
 \includegraphics[bb=0 0 300 200, angle=0, scale=1.2]{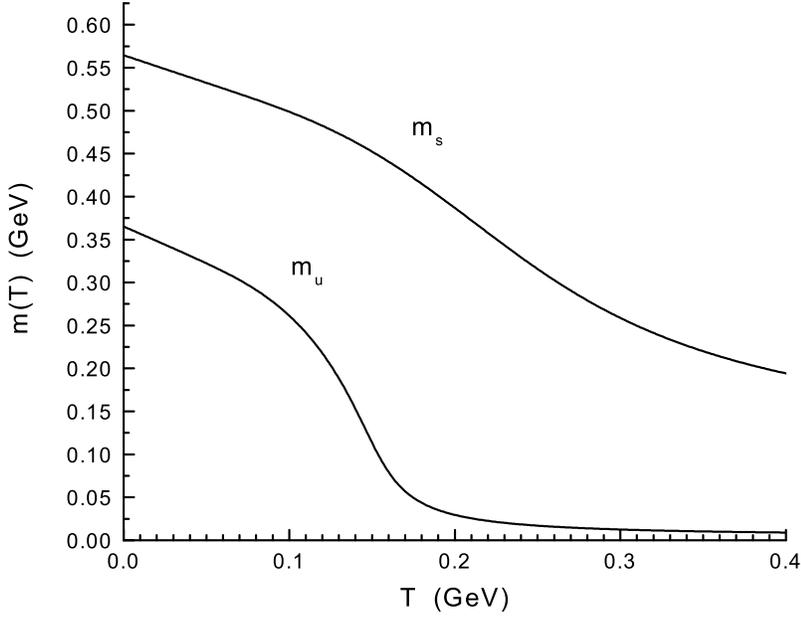}%
 \caption{Temperature dependent constituent mass values, $m_u(T)$ and $m_s(T)$,
 calculated using Eq. (3.2) are shown. Here $m_u^0=0.0055$ GeV,
 $m_s^0=0.120$ GeV, and $G(T)=5.691[1-0.17(T/T_c)]$, if we use
 Klevansky's notation [15]. (The value of $G_S$ of Eq. (1.1) is
 twice the value of $G$ used in Ref. [15]).}
 \end{figure}

\section{random phase approximation calculations for meson masses at finite temperature}

The analysis of the diagrams of Fig. 3b gives rise to a set of
equations for various vertex functions. These equations are of the
form of relativistic random-phase-approximation equations. The
derivation of these equations for pseudoscalar mesons is given in
Ref. [6], where we discuss the equations for pionic, kaonic and
eta mesons. The equations for the eta mesons are the most
complicated, since we consider singlet-octet mixing as well as
pseudoscalar--axial-vector mixing. In that case there are eight
vertex functions to consider, $\Gamma_{P,0}^{+-}$,
$\Gamma_{A,0}^{+-}$, $\Gamma_{P,8}^{+-}$, $\Gamma_{A,8}^{+-}$,
$\Gamma_{P,0}^{-+}$, $\Gamma_{A,0}^{-+}$, $\Gamma_{P,8}^{-+}$,
$\Gamma_{A,8}^{-+}$, where $P$ refers to the $\gamma_5$ vertex and
$A$ refers to the $\gamma_0\gamma_5$ vertex, which mixes with the
$\gamma_5$ vertex. Corresponding to the eight vertex functions one
may define eight wave function amplitudes [6].

The simplest example of our RPA calculations is that of the $a_0$
mesons, where there are only two vertex functions $\Gamma_S^{+-}$
and $\Gamma_S^{-+}$ to be calculated [4]. Associated with
$\Gamma_S^{+-}$ and $\Gamma_S^{-+}$ are two wave functions
$\phi_S^+$ and $\phi_S^-$, which are the large and small
components respectively. In vacuum one has coupled equations for
these wave functions. \be 2E_u(k)\phi^+(k)+\int\!
d\kp\,[H_C(k,\kp)+H_{NJL}(k,\kp)]\phi^+(\kp)\\\nonumber+\int\!
d\kp \,H_{NJL}(k,\kp)\phi^-(\kp)=P^0\phi^+(k)\,,\ee \be
-2E_u(k)\phi^-(k)-\int\!
d\kp\,[H_C(k,\kp)+H_{NJL}(k,\kp)]\phi^-(\kp)\\\nonumber-\int\!
d\kp \,H_{NJL}(k,\kp)\phi^+(\kp)=P^0\phi^-(k)\,,\ee where
$E_u(k)=[\vec k\,{}^2+m_u^2]^{1/2}$, \be
H_C(k,\kp)=-\frac1{(2\pi)^2}\frac{[2V_0^C(k,\kp)k^2\kpsq+m_u^2k\kp
V_1^C(k,\kp)]}{E_u(k)E_u(\kp)}\,,\ee and \be
H_{NJL}=\frac{8N_c}{(2\pi)^2}\frac{\ksq\kpsq
G_{a_0}e^{-\ksq/2\alpha^2}e^{-\kpsq/2\alpha^2}}{E_u(k)E_u(\kp)}\,.\ee
In Eq. (4.4), $G_{a_0}$ is the effective coupling constant for the
$a_0$ mesons, which depends upon the values of $G_S$, $G_D$ and
the vacuum condensates. These relations for the various coupling
constants, $G_{a_0}$, $G_\pi$, $G_K$, $G_{f_0}$, $G_{K_0^*}$, etc.
may be found in Ref. [16]. In Eq. (4.3) we have introduced \be
V_l^C(k,\kp)=\int_{-1}^1\!dx\, P_l(x)V^C(\vec k-\vec
k{}^\prime)\,.\ee Here, $x=\mbox{cos}\theta$ and $P_l(x)$ is a
Legendre function. The terms exp[$-\ksq/2\alpha^2$] and
exp[$-\kpsq/2\alpha^2$] are regulators with $\alpha=0.605$ GeV.

In order to solve these equations at finite temperature, we
replace $m_u$, $G_{a_0}$ and $\mu_0$ by $m_u(T)$, $G_{a_0}(T)$ and
$\mu(T)$. The values of $m(T)$ are given in Fig. 5, and we recall
that $\mu(T)=\mu_0/[1-0.7(T/T_c)^2]$. In the RPA, the solutions of
Eqs. (4.1) and (4.2) come in pairs. For a state of energy $P_i^0$
there is another state with energy $-P_i^0$. Since the RPA
Hamiltonian is not Hermitian, it is possible to obtain imaginary
values for the energy. That is a signal of the instability of the
ground state of the theory and requires that the problem be
reformulated to obtain a stable ground state. This problem does
not arise in the calculations reported in this work. In
particular, the use of the temperature-dependent values of
$G_\pi(T)$ avoids the appearance of pion condensation in the
formalism.

\section{results of numerical calculations of meson masses at \\finite temperature}

As noted earlier, the RPA equations for the $a_0$ mesons are given
in Ref. [4] and those for the $\pi$, $K$ and $\eta$ mesons are
given in Ref. [6]. The equations needed in the study of the $f_0$
mesons are to be found in Ref. [5], while the RPA equations for
the study of the $K_0^*$ mesons are to be found in the Appendix of
Ref. [3]. In Figs. 6-10 we present our results for the $\pi$, $K$,
$a_0$, $f_0$ and $K_0^*$ mesons. The values of the coupling
constants used are given in the figure captions. The reduction of
the number of bound states with increasing temperature can be
understood by noting that for the $\pi$ and $a_0$ mesons the
continuum of the model lies above $V_{max}(T)+2m_u(T)$, while for
the $K$ and $K_0^*$ mesons the continuum lies above
$V_{max}(T)+m_u(T)+m_s(T)$. The situation is more complex in the
case of the $f_0$ mesons which contain both $s\bar s$, $u\bar u$
and $d\bar d$ components. The bound states lie below the $s\bar s$
continuum which begins at $V_{max}(T)+2m_s(T)$. However, we note
that the absence of bound states at $T=T_c$ for all the mesons
considered here is due to the reduction of the value of the
confining potential and of the constituent quark masses.

It is of interest to note that the mass values of the $a_0$ and
$f_0$ mesons tend toward degeneracy with the pion as $T\rightarrow
T_c$\,. However, the mesons disassociate before a greater
degeneracy is achieved. That is in contrast to the results
obtained in the SU(2) formalism considered by Hatsuda and Kunihiro
[16]. Since these authors do not include a model of the
confinement-deconfinement transition, they are able to see the
approximate degeneracy of the sigma meson and the pion with
increasing temperature. It is also worth noting that in the SU(3)
formalism the sigma meson is replaced by the $f_0 (980)$ as the
chiral partner of the pion.

 \begin{figure}
 \includegraphics[bb=0 0 440 670, angle=0, scale=0.5]{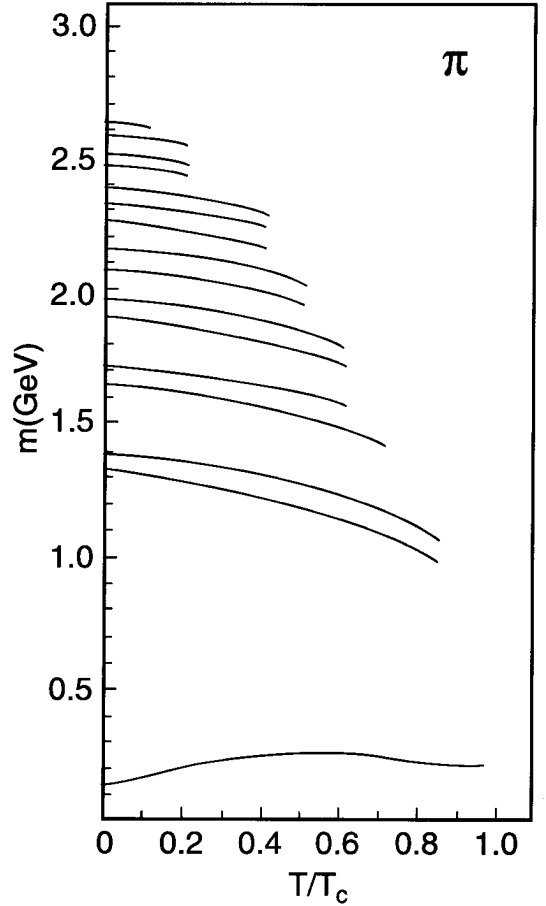}%
 \caption{The mass values of the pionic states calculated in this work
 with $G_\pi(T)=13.49[1-0.17\,T/T_c]$ GeV, $G_V(T)=11.46[1-0.17\,T/T_c]$ GeV,
 and the quark mass values given in Fig. 5. The value of the pion mass is
 0.223 GeV at $T/T_c=0.90$, where $m_u(T)=0.102$ GeV and $m_s(T)=0.449$ GeV. The
 pion is bound up to $T/T_c=0.94$, but is absent beyond that value.}
 \end{figure}

 \begin{figure}
 \includegraphics[bb=0 0 440 670, angle=0, scale=0.5]{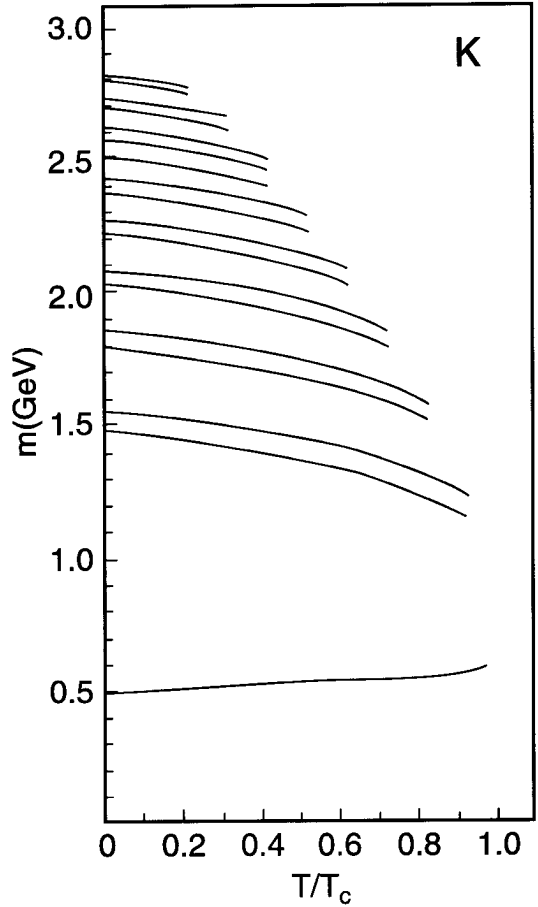}%
 \caption{Mass values of kaonic states calculated with $G_K(T)=
 13.07[1-0.17\,T/T_c]$ GeV, $G_V(T)=11.46[1-0.17\,T/T_c]$ GeV, and the quark
 mass values given in Fig. 5. The value of the kaon mass is 0.598 GeV at
 $T/T_c=0.95$, where $m_u(T)=0.075$ GeV and $m_s(T)=0.439$ GeV.}
 \end{figure}

 \begin{figure}
 \includegraphics[bb=0 0 440 670, angle=0, scale=0.5]{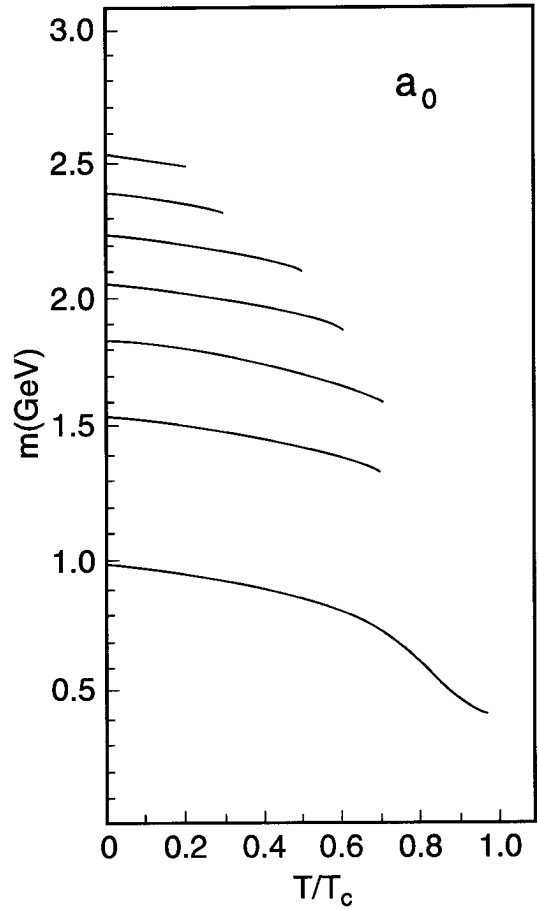}%
 \caption{Mass values for the $a_0$ mesons calculated with $G_{a_0}(T)=
 13.1[1-0.17\,T/T_c]$ GeV, and the quark mass values given in Fig. 5.
 The value of the $a_0$ mass at $T/T_c=0.95$ is 0.416 GeV.}
 \end{figure}

 \begin{figure}
 \includegraphics[bb=0 0 440 670, angle=0, scale=0.5]{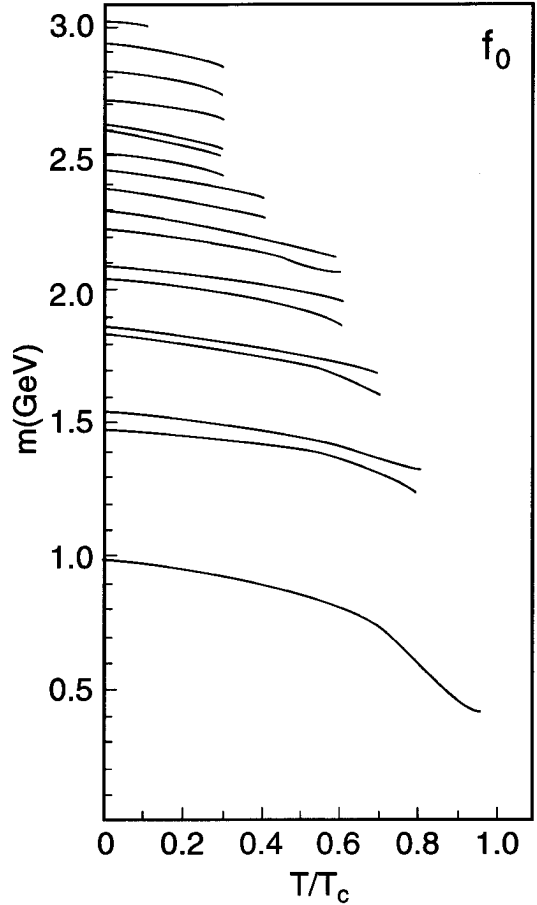}%
 \caption{Mass values of the $f_0$ mesons calculated with
 $G_{00}(T)=14.25[1-0.17\,T/T_c]$ GeV, $G_{88}(T)=10.65[1-0.17\,T/T_c]$ GeV,
 $G_{08}(T)=0.495[1-0.17\,T/T_c]$ GeV, and $G_{80}(T)=G_{08}(T)$ in a singlet-octet
 representation. The quark mass values used are shown in Fig. 5. The
 $f_0$ has a mass of 0.400 GeV at $T/T_c=0.95$.}
 \end{figure}

 \begin{figure}
 \includegraphics[bb=0 0 440 670, angle=0, scale=0.5]{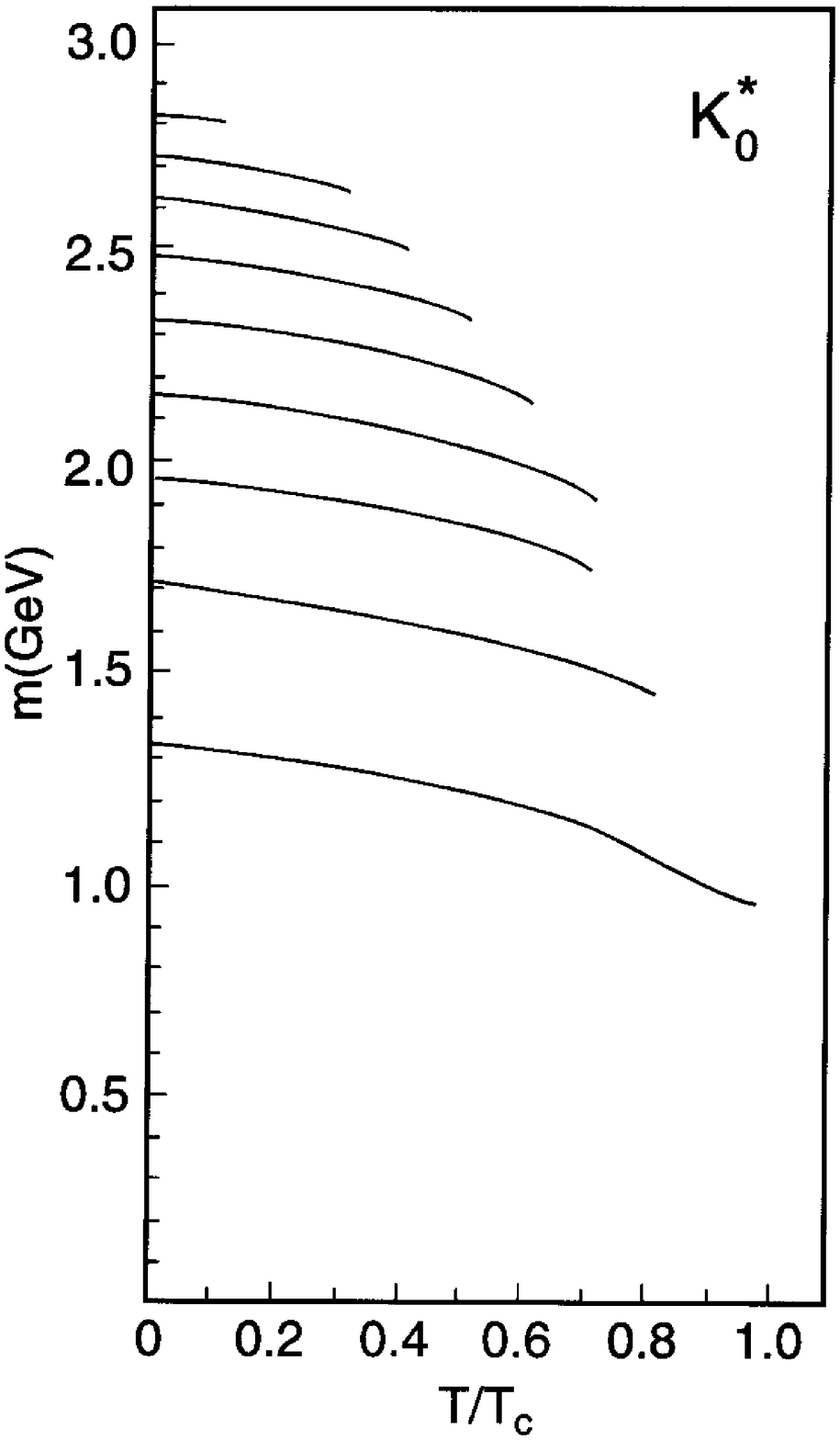}%
 \caption{Mass values obtained for the $K_0^*$ mesons with $G_{K_0^*}(T)=
 10.25[1-0.17\,T/T_c]$ GeV, and the quark mass values shown in Fig. 5.}
 \end{figure}

In order to demonstrate the interplay of chiral symmetry
restoration and dissolution of our meson states, we have performed
calculations in which the quark masses are unchanged from their
value at $T=0$. For the pion, with $m_u=m_d=0.364$ GeV and for
$T/T_c=0.95$, we find bound states at 0.530, 1.242 and 1.305 GeV.
If we also consider the values of the coupling constants and quark
masses fixed at their $T=0$ values, bound states are found at
0.102, 2.248 and 1.298 GeV when $T/T_c=0.95$. A similar analysis
for the kaon states yields bound states at $T=T_c$ of 0.738, 1.395
and 1.444 GeV, if we put $m_u=0.364$ GeV and $m_s=0.565$ GeV. If,
in addition, we neglect the temperature dependence of the coupling
constants, we have bound kaon states at 0.482, 1.440 and 1.439
GeV.

In the case of the $a_0$ states, putting $m_u=m_d=0.364$ GeV
yields a single state at 0.960 GeV at $T=T_c$\,, if we neglect the
temperature dependence of the coupling constant. If we maintain
the temperature dependence of the coupling constant, we find a
single state at 1.067 GeV, if the quark masses are kept at their
$T=0$ values.

From this analysis, we see that the reduction of the quark masses
with increasing temperature, which represents a partial
restoration of chiral symmetry, is an essential feature of our
model. In the past, lattice simulations of QCD have indicated that
deconfinement and restoration of chiral symmetry take place at the
same temperature. Since our Lagrangian contains current quark
masses for the up, down and strange quarks, chiral symmetry is not
completely restored at the higher temperatures in our model.
However, the model does exhibit quite significant reductions of
the up and down constituent quark masses for $T\gtrsim T_c$ (See
Fig. 5), so that deconfinement is here associated with a
significant restoration of chiral symmetry.

\section{discussion}

In recent years we have seen extensive applications of the NJL
model in the study of matter at high density [17-21]. There is
great interest in the diquark condensates and color
superconductivity predicted by the NJL model and closely related
models that are based upon instanton dynamics. It is noted by
workers in this field that the NJL model does not describe
confinement, with the consequence that one can not present a
proper description of the hadronic phase that exists at the
smaller values of the temperature and density in the QCD phase
diagram. Thus, the attitude adopted is that, if one works in the
deconfined phase, the NJL model may provide a satisfactory
description of the quark interaction. In the present work we have
modified the NJL model so that we can describe light mesons and
their radial excitations, as well as the confinement-deconfinement
transition at $T=T_c$\,. In an earlier work we studied the
confinement-deconfinement transition for finite matter density at
$T=0$ [13]. (A more comprehensive study would include both finite
temperature and finite density.) As in the present work, in which
we introduced a temperature-dependent coupling constants, we used
density-dependent coupling constants in Ref. [13]. If such
dependence exists, it would have important consequences for the
study of dense matter using the NJL model.

One interesting feature of our results is that both the lowest
$a_0$ and $f_0$ states move toward degeneracy with the pion as the
temperature is increased. However, the system is deconfined before
such degeneracy can be exhibited.

We stress that the restoration of chiral symmetry is intimately
connected with the dissolution of our meson states at $T=T_c$\,.
As we saw in the discussion toward the end of Section V, the
various mesons studied here still have bound states at $T=T_c$\,,
if only the temperature dependence of the confining field is
included in the model.

The behavior of charmonium across the deconfinement transition has
recently been studied using lattice simulations of QCD [22]. The
authors of Ref. [22] point out that, unlike the case of light
mesons, charmonia may exist as bound states even after the
deconfinement transition. They state:``Our studies support the
sequential pattern for charmonium  dissolution obtained from
potential model studies, where the broader bound states (the
scalar and axial vector channels) dissolve before the pseudoscalar
and vector channels [23]. The pseudoscalar and vector channels are
seen to survive as bound states still at 1.25$T_c$ and probably
dissolve after 1.5$T_c$\,."

\renewcommand{\theequation}%
    {\Alph{section}\arabic{equation}}
\appendix*
\section{}

Since the introduction of temperature-dependent coupling constants
for the NJL model is a novel feature of our work, we provide
arguments in this Appendix to justify their introduction. We make
reference to Fig. 1.3 of Ref. [24]. That figure shows the behavior
of the ratio $\epsilon/T^4$ and $3P/T^4$ for the pure gauge sector
of QCD. Here $\epsilon$ is the energy density and $P$ is the
pressure. Ideal gas behavior implies $\epsilon=3P$. The values of
$\epsilon/T^4$ and $3P/T^4$ are compared to the value
$\epsilon_{SB}/T^4=8\pi^2/15$ for an ideal gluon gas. It may be
seen from the figure that at $T=3T_c$ there are still significant
differences from the ideal gluon gas result. Deviations from ideal
gas behavior become progressively smaller with increasing $T/T_c$
and could be considered to be relatively unimportant for
$T/T_c>5$.

To provide evidence for temperature-dependent coupling constants,
we discuss the calculation of hadronic current correlators in the
deconfined phase. The procedure we adopt is based upon the
real-time finite-temperature formalism, in which the imaginary
part of the polarization function may be calculated. Then, the
real part of the function is obtained using a dispersion relation.
The result we need for this work has been already given in the
work of Kobes and Semenoff [25]. (In Ref. [25] the quark momentum
is $k$ and the antiquark momentum is $k-P$. We will adopt that
notation in this section for ease of reference to the results
presented in Ref. [25].) With reference to Eq. (5.4) of Ref. [25],
we write the imaginary part of the scalar polarization function as
\be \mbox{Im}\,J^S(P^2,
T)=\frac12(2N_c)\beta_S\,\epsilon(p^0)\mytint ke^{-\vec
k\,{}^2/\alpha^2}\left(\frac{2\pi}{2E_1(k)2E_2(k)}\right)\\\nonumber
\{(1-n_1(k)-n_2(k))
\delta(p^0-E_1(k)-E_2(k))\\\nonumber-(n_1(k)-n_2(k))
\delta(p^0+E_1(k)-E_2(k))\\\nonumber-(n_2(k)-n_1(k))
\delta(p^0-E_1(k)+E_2(k))\\\nonumber-(1-n_1(k)-n_2(k))
\delta(p^0+E_1(k)+E_2(k))\}\,.\ee Here, $E_1(k)=[\vec
k\,{}^2+m_1^2(T)]^{1/2}$. Relative to Eq. (5.4) of Ref. [25], we
have changed the sign, removed a factor of $g^2$ and have included
a statistical factor of $2N_c$, where the factor of 2 arises from
the flavor trace. In addition, we have included a Gaussian
regulator, $\exp[-\vec k\,{}^2/\alpha^2]$, with $\alpha=0.605$
GeV, which is the same as that used in most of our applications of
the NJL model in the calculation of meson properties. We also note
that \be n_1(k)=\frac1{e^{\,\beta E_1(k)}+1}\,,\ee and \be
n_2(k)=\frac1{e^{\,\beta E_2(k)}+1}\,.\ee For the calculation of
the imaginary part of the polarization function, we may put
$\ksq=m_1^2(T)$ and $(k-p)^2=m_2^2(T)$, since in that calculation
the quark and antiquark are on-mass-shell. We will first remark
upon the calculation of scalar correlators [25]. In that case, the
factor $\beta_S$ in Eq. (A1) arises from a trace involving Dirac
matrices, such that
\be \beta_S&=&-\mbox{Tr}[(\slr k+m_1)(\slr k-\slr P+m_2)]\\
&=&2P^2-2(m_1+m_2)^2\,,\ee where $m_1$ and $m_2$ depend upon
temperature. In the frame where $\vec P=0$, and in the case
$m_1=m_2$, we have $\beta_S=2P_0^2(1-{4m^2}/{P_0^2})$. For the
scalar case, with $m_1=m_2$, we find \be \mbox{Im}\,J^S(P^2,
T)=\frac{N_cP_0^2}{4\pi}\left(1-\frac{4m^2}{P_0^2}\right)^{3/2}
e^{-\vec k\,{}^2/\alpha^2}[1-2n_1(k)]\,,\ee where \be \vec
k\,{}^2=\frac{P_0^2}4-m^2(T)\,.\ee

We may evaluate Eq. (2.8) for $m(T)=m_u(T)=m_d(T)$ and define
$\mbox{Im}\,J_u^S(P^2, T)$. Then we put $m(T)=m_s(T)$, we define
$\mbox{Im}\,J_s^S(P^2, T)$. These two functions are needed for a
calculation of the scalar-isoscalar correlator. The real parts of
the functions $J_u^S(P^2, T)$ and $J_s^S(P^2, T)$ may be obtained
using a dispersion relation, as noted earlier.

For pseudoscalar mesons, we replace $\beta_S$ by
\be \beta_P&=&-\mbox{Tr}[i\gamma_5(\slr k+m_1)i\gamma_5(\slr k-\slr P+m_2)]\\
&=&2P^2-2(m_1-m_2)^2\,,\ee which for $m_1=m_2$ is $\beta_P=2P_0^2$
in the frame where $\vec P=0$. We find, for the $\pi$ mesons, \be
\mbox{Im}\,J^P(P^2,T)=\frac{N_cP_0^2}{4\pi}\left(1-\frac{4m(T)^2}{P_0^2}\right)^{1/2}
e^{-\vec k\,{}^2/\alpha^2}[1-2n_1(k)]\,,\ee where $ \vec
k\,{}^2={P_0^2}/4-m_u^2(T)$, as above. Thus, we see that, relative
to the scalar case, the phase space factor has an exponent of 1/2
corresponding to a \textit{s}-wave amplitude, rather than the
\textit{p}-wave amplitude of scalar mesons. For the scalars, the
exponent of the phase-space factor is 3/2, as seen in Eq. (A6).

For a study of vector mesons we consider \be
\beta_{\mu\nu}^V=\mbox{Tr}[\gamma_\mu(\slr k+m_1)\gamma_\nu(\slr
k-\slr P+m_2)]\,,\ee and calculate \be
g^{\mu\nu}\beta_{\mu\nu}^V=4[P^2-m_1^2-m_2^2+4m_1m_2]\,,\ee which,
in the equal-mass case, is equal to $4P_0^2+8m^2(T)$, when
$m_1=m_2$ and $\vec P=0$. Note that for the elevated temperatures
considered in this work $m_u(T)=m_d(T)$ is quite small, so that
$4P_0^2+8m_u^2(T)$ can be approximated by $4P_0^2$ when we
consider the $\rho$ meson.

We now consider the calculation of temperature-dependent hadronic
current correlation functions. The general form of the correlator
is a transform of a time-ordered product of currents, \be C(P^2,
T)=i\int d^4xe^{\,ip\cdot x}<<\mbox T(j(x)j(0)>>\,,\ee where the
double bracket is a reminder that we are considering the finite
temperature case.

For the study of pseudoscalar states, we may consider currents of
the form $j_{P,\,i}(x)=\bar q(x)i\gamma_5\lambda^iq(x)$, where, in
the case of the $\pi$ mesons, $i=1,2,$ and 3. For the study of
pseudoscalar-isoscalar mesons, we again introduce
$j_{P,\,i}(x)=\bar q(x)\lambda^iq(x)$, but here $i=0$ for the
flavor-singlet current and $i=8$ for the flavor-octet current.

In the case of the $\pi$ mesons, the correlator may be expressed
in terms of the basic vacuum polarization function of the NJL
model, $J_P(P^2, T)$. Thus, \be C_\pi(P^2,T)=J_P(P^2,
T)\frac1{1-G_\pi(T)J_P(P^2, T)}\,,\ee where $G_\pi(T)$ is the
coupling constant appropriate for our study of the $\pi$ mesons.
We have found $G_\pi(0)=13.49$\gev{-2} by fitting the pion mass in
a calculation made at $T=0$, with $m_u=m_d=0.364$ GeV.

The calculation of the correlator for pseudoscalar-isoscalar
states is more complex, since there are both flavor-singlet and
flavor-octet states to consider. We may define polarization
functions for $u$, $d$ and $s$ quarks: $J_u(P^2, T)$, $J_d(P^2,
T)$ and $J_s(P^2, T)$. In terms of these polarization functions we
may then define \be J_{00}(P^2, T)=\frac23[J_u(P^2, T)+J_d(P^2,
T)+J_s(P^2, T)]\,,\ee \be J_{08}(P^2, T)=\frac{\sqrt2}3[J_u(P^2,
T)+J_d(P^2, T)-2J_s(P^2, T)]\,,\ee and \be J_{88}(P^2,
T)=\frac13[J_u(P^2, T)+J_d(P^2, T)+4J_s(P^2, T)]\,.\ee We also
introduce the matrices \be J(P^2,
T)=\left[\begin{array}{cc}J_{00}(P^2, T)&J_{08}(P^2,
T)\\J_{80}(P^2, T)&J_{88}(P^2, T)\end{array}\right]\,,\ee \be
G(T)=\left[\begin{array}{cc}G_{00}(T)&G_{08}(T)\\G_{80}(T)&G_{88}
(T)\end{array}\right]\,,\ee and \be C(P^2,
T)=\left[\begin{array}{cc}C_{00}(P^2, T)&C_{08}(P^2,
T)\\C_{80}(P^2, T)&C_{88}(P^2, T)\end{array}\right]\,.\ee We then
write the matrix relation \be C(P^2, T)=J(P^2, T)[1-G(T)J(P^2,
T)]^{-1}\,.\ee

The use of our energy-dependent coupling constants is meant to be
consistent with the approach to asymptotic freedom at high
temperature. In order to understand this feature in our model, we
can calculate the correlator with \emph{constant} values of
$G_{00}$, $G_{88}$ and $G_{08}$ and also with
$G_{00}(T)=G_{00}[1-0.17\,T/T_c]$, etc. (In this work we use
$G_{00}=8.09$\gev{-2}, $G_{88}=13.02$\gev{-2} and
$G_{08}=-0.4953$\gev{-2}.)

We now consider the values of $\mbox{Im}\,C_{88}(P^2)$ for
$T/T_c=4.0$. In Fig. 11 we show the values of
$\mbox{Im}\,C_{88}(P^2)$ calculated in our model with
temperature-dependent coupling parameters as a dashed line. The
dotted line shows the values of the correlator for
$G_{00}=G_{88}=G_{08}=0$, while the solid line shows the values
when the coupling parameters are kept at their values at $T=0$. We
see that we have resonant behavior in the case the parameters are
temperature independent.

In Fig. 12 we show similar results for $T/T_c=5.88$. Here the
temperature-dependent coupling constants are equal to zero, so
that the lines corresponding to the dashed and dotted lines of
Fig. 11 coincide. The solid line again shows resonant behavior at
a value of $T/T_c$ where we expect only weak interactions
associated with asymptotic freedom. We conclude that the model
with constant values of the coupling parameters yields
unacceptable results, while our model, which has
temperature-dependent coupling parameters, behaves as one may
expect, when the results of lattice simulations of QCD
thermodynamics are taken into account.

\begin{figure}
\includegraphics[bb=0 0 300 200, angle=0, scale=1.2]{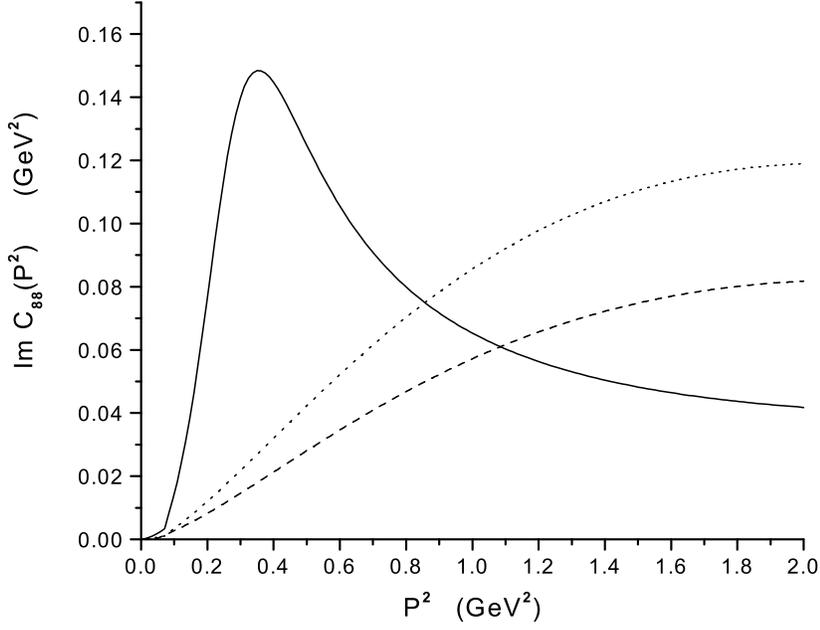}%
\caption{The imaginary part of the correlator $C_{88}(P^2)$ is
shown for $T/T_c=4.0$. The dashed line is the result for the
temperature-dependent coupling parameters of our model, while the
solid line represents the results for coupling parameters kept at
their $T=0$ values. The dotted line shows the values of the
correlator when the coupling parameters are set equal to zero.}
\end{figure}

\begin{figure}
\includegraphics[bb=0 0 300 200, angle=0, scale=1.2]{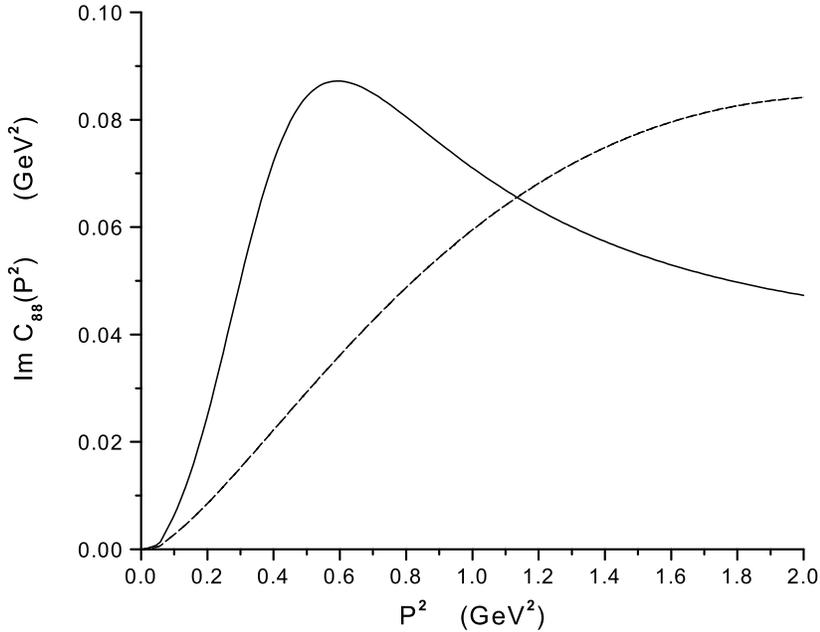}%
\caption{The imaginary part of the correlator $C_{88}(P^2)$ is
shown for $T/T_c=5.88$. [See caption to Fig. 11.] Here the dashed
and dotted lines of Fig. 11 coincide.}
\end{figure}




\end{document}